\documentclass[journal]{journal}

\usepackage{cite}
\usepackage{graphicx}
\usepackage[cmex10]{amsmath}
\usepackage{amssymb}
\usepackage{array}
\usepackage{caption}
\usepackage[none]{hyphenat}
\usepackage[font=footnotesize]{caption}
\usepackage[justification=centering]{caption} 
\usepackage{textcmds}
\newcolumntype{C}[1]{>{\raggedright\let\newline\\\arraybackslash\hspace{0pt}}m{#1}}

\begin{document}

\title{Q-Map: Clinical Concept Mining \\from Clinical Documents}

\author{Sheikh Shams Azam,~Manoj Raju,~Venkatesh Pagidimarri,~Vamsi Kasivajjala
\thanks{M. Raju (Senior Data Scientist), V. Pagidimarri (General Manager) and V. Kasivajjala (Senior Vice President) are with the Data Science Team of Enlightiks, Practo Technologies Pvt. Ltd.,  Bangalore, Karnataka, 560076 (e-mail: \{manoj.raju, venkatesh.p, vamsi\}@practo.com).}
\thanks{S. S. Azam (Data Scientist) is with the Data Science Team of Enlightiks, Practo Technologies Pvt. Ltd.,  Bangalore, Karnataka, 560076 (e-mail: s.shams.sam@gmail.com, see https://machinelearningmedium.com/about/).}
}

\markboth{Journal of \LaTeX\ Class Files,~Vol.~6, No.~1, January~2007}%
{Shell \MakeLowercase{\textit{et al.}}: Bare Demo of IEEEtran.cls for Journals}

\maketitle
\thispagestyle{empty}

\begin{abstract}

Over the past decade, there has been a steep rise in the data-driven analysis in major areas of medicine, such as clinical decision support system, survival analysis, patient similarity analysis, image analytics etc. Most of the data in the field are well-structured and available in numerical or categorical formats which can be used for experiments directly. But on the opposite end of the spectrum, there exists a wide expanse of data that is intractable for direct analysis owing to its unstructured nature which can be found in the form of discharge summaries, clinical notes, procedural notes which are in human written narrative format and neither have any relational model nor any standard grammatical structure. An important step in the utilization of these texts for such studies is to transform and process the data to retrieve structured information from the haystack of irrelevant data using information retrieval and data mining techniques. To address this problem, the authors present Q-Map in this paper, which is a simple yet robust system that can sift through massive datasets with unregulated formats to retrieve structured information aggressively and efficiently. It is backed by an effective mining technique which is based on a string matching algorithm that is indexed on curated knowledge sources, that is both fast and configurable. The authors also briefly examine its comparative performance with MetaMap, one of the most reputed tools for medical concepts retrieval and present the advantages the former displays over the latter.

\end{abstract}

\begin{IEEEkeywords}
Information Retrieval (IR), Unified Medical Language System (UMLS), Syntax Based Analysis, Natural Language Processing (NLP), Medical Informatics.
\end{IEEEkeywords}

\IEEEpeerreviewmaketitle

\section{Introduction}

\IEEEPARstart{T}{here} are various traditional approaches for retrieving structured information from clinical free text, which are generally based on term importance metrics from text corpora and are not very efficient in determining the similarity between relevant formats and encountered texts, even if there are small variations between the sequences to be matched. These methods suffer from several other issues, such as data sparseness, high dimensionality, out of vocabulary (OOV) words, ambiguous words in the corpus. Other hardware limitations are also encountered in processing the data, such as memory usage and time complexity.

Q-Map tries to circumvent these issues while retaining a robust retrieval mechanism for efficient discovery and annotation of medical concepts from the clinical free text. We try to solve issues of concept aggregation that are irregular in Unified Medical Language System (UMLS) \cite{bodenreider} Knowledge Sources \cite{lindberg} using data-driven approaches. We also introduce a rule-based phrase sense detection algorithm for marking the negated concepts in a text.

The algorithm aims to find the phrases which are medically relevant in a single step parse. The next stage then eliminates the phrases that seem to be negated in the context. For example, if the text contains \qq{Patient does not have diabetes}, then the term \qq{diabetes} should not be detected positive even though it has a medical association. Together with spell correction, negation detection and phrase detection using the UMLS Metathesaurus \cite{schuyler}, the algorithm provides a very fast way of parsing large text dumps. 

The rest of the paper is structured as follows. Section II describes the background related to the work. Section III briefly explains related works in the field. In section IV, we dive into the data, methodology, algorithms, and architecture of the system. Section V presents results of comparison between Q-Map and MetaMap \cite{aronson} with examples. Conclusions are summarized in section VI. Finally, we end with acknowledgement and references.


\section{Background}

\subsection{Clinical Documents}

Clinical Documents are the end result of the clinical documentation, which can be defined as the generation of digital or analog medical record detailing the course of treatment, clinical trial or tests undergone by an out-patient or an in-patient. There are different types of clinical documents namely, discharge summary, progress notes (summarises the diagnostic findings and patient status on a daily basis), surgical reports (surgical procedures are noted as a written operative report), radiology reports etc. Most of these clinical documents are in free text format where the detailed reports are narratives by physicians, care-providers, or specialists. These clinical documents serve as input to the Q-Map system after it is trained to extract systematic data from them.

\subsection{UMLS Metathesaurus}

The main requirement for building an Information Retrieval (IR) system is to have an exhaustive knowledge source. The UMLS is a collection of concise but detailed medical concepts from many controlled vocabularies in biomedical sciences. UMLS Metathesaurus is the base knowledge source of UMLS which is a database consisting of over 1 million biomedical concepts and 5 million concept names, which are collected and periodically updated from over 100 different controlled vocabularies like, ICD-10 \cite{icd}, MeSH \cite{lipscomb}, SNOMED CT \cite{donnelly}, LOINC \cite{mcdonald}, PubMed \cite{pubmed}, PMC \cite{pmc}, RxNorm \cite{liu} etc. UMLS Metathesaurus organises all these data from different sources under different Concept Unique Identifiers (CUI), which effectively helps in organising synonymous terms. UMLS also creates another hierarchy by grouping similar CUIs under semantic types. Semantic types are identifiers denoting the class of concept strings. UMLS has 133 different semantic types, such as finding, disease and syndrome, injury or poisoning etc. UMLS Metathesaurus configured using options, such as CUI clustering (explained in IV-A) and semantic type filtering (explained in IV-D), forms the knowledge base of the system we developed in this work.

\section{Related Work}

Many different systems have been developed in the past for the purpose of text mining. Few of such systems are Metaphrase \cite{tuttle}, CLARIT \cite{evans}, MetaMap, cTAKES \cite{savova}. These systems have been employed in various different applications and have had a varying degree of success based on how well the data is structured.  One of the most reputed tool for medical text mining currently is MetaMap.  MetaMap builds on top of UMLS Metathesaurus and SPECIALIST lexicon \cite{browne}, but suffers from the issue of massive complexity, and hence debugging and additions to the system is a challenging task. 

Q-Map tries to achieve the same objective with a much simpler approach with better flexibility in the system while producing a comparable performance with added benefits of speed and handling huge amounts of datasets. We present the Q-Map architecture and its future scope and present a comparative study with MetaMap. Our original contribution includes the design and architecture of the system and the ensemble of algorithms used with various modifications explained in the following sections, that help put together a stack efficient in clinical concept retrieval. We also try to remove the discrepancies in UMLS knowledge sources using data driven approaches explained in next section.

\section{Procedure and Architecture}

\subsection{CUI Clustering/Bagging}

Although the UMLS Metathesaurus is a comprehensive collection of medical concepts, it has some flaws in data organization. Concepts which have the same meaning but comprise of different root words are often bagged under different CUIs. We try to remove such discrepancies by clubbing all the similar terms under unique identifiers using natural language processing (NLP) techniques.

This bagging algorithm can be configured to define how aggressively the concepts need to be clustered. It is based on thresholds of concept mapping metrics which is based on edit distance. For example, bagging can be performed with stemming of concept terms which would be more aggressive in mapping the terms, or it can be performed without stemming. Similarly, there are options to remove stopwords and consider the word order. After this process, a masking table is produced which maps the old CUIs to the new ones. For example, \qq{diabetes} occurs under the following CUIs - C0011847, C0011849, C0011860 in Metathesaurus, but after the bagging process, all the terms under the 3 CUIs will be clustered together under one CUI. Our algorithm also takes care of transitive nature of concept matching during this clustering process. For example, \qq{diabetes} occurs under CUIs - C0011847, C0011849 and \qq{DM} (Diabetes Mellitus) occurs under CUIs - C0011849, C3250443, so effectively all the terms under these CUIs are grouped together under a single CUI. 

This masking process gives promising results in the subsequent classification tasks fed with features retrieved from Q-Map.

\subsection{Aho-Corasick Algorithm}

Aho-Corasick algorithm \cite{aho} is a string search algorithm similar to the implementation of a trie but is optimized by maintaining failure states in case of an unsuccessful character match. It is a finite state machine (FSM) with characters as its transition states and failure loops as its fall-back states. This dictionary matching algorithm locates the elements of a finite set of strings in the dictionary within a text document. The complexity of the search is linearly proportional to the length of the string from which the medical concepts are being mined. The failure states help in fast transition between states whenever a character mismatch is encountered. This efficient failure transition prevents the algorithm from backtracking to the root to restart the matching process.

  \begin{figure*}[thpb]
      \centering
      \framebox{\includegraphics[scale=0.35]{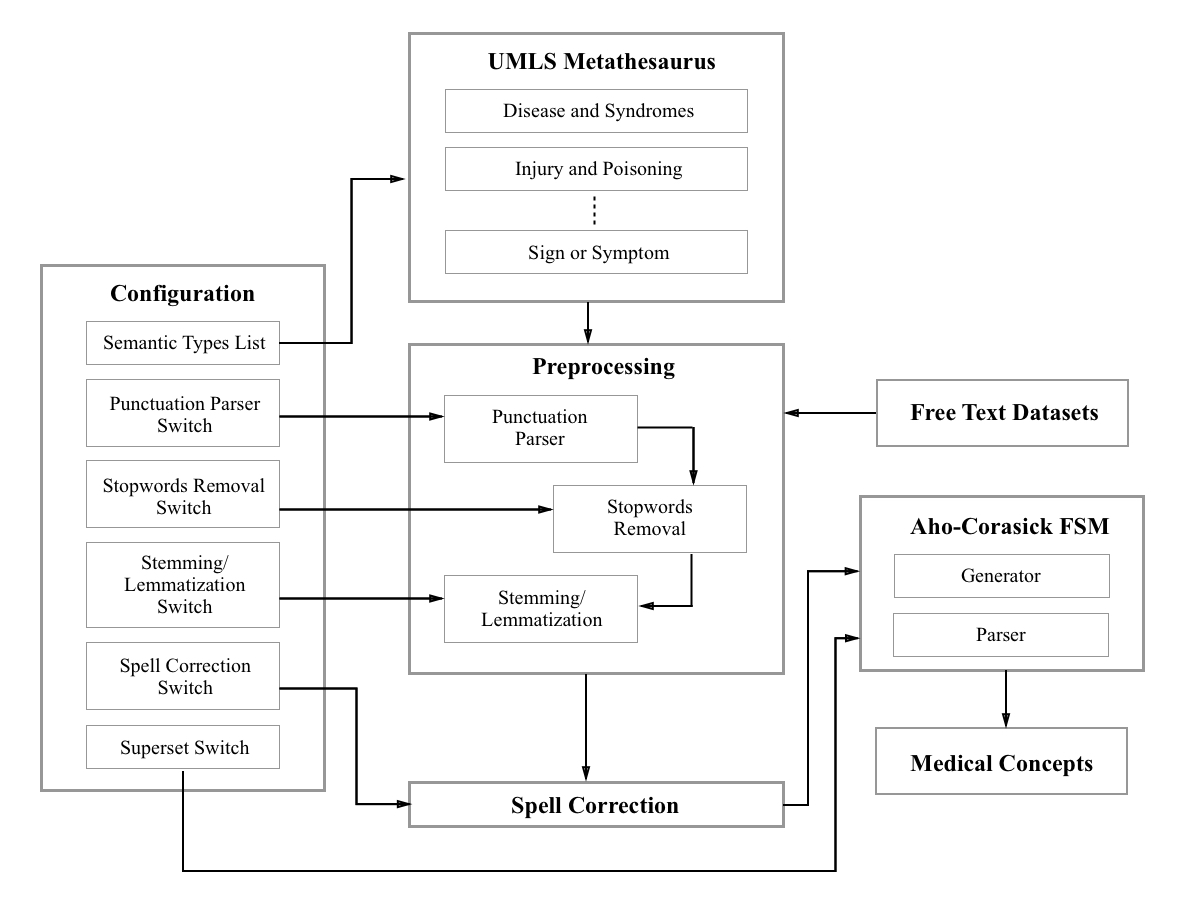}}
      \caption{Q-Map Architecture Flowchart}
      \label{figurelabel}
   \end{figure*}

\subsection{Spell Correction}

Edit distance is one of the most popular ways of determining string similarity. An edit in the edit distance can be one of the following operations: deletion (removal of a character from the string), addition (add a new character to the string), substitution (replace a character with another in the string) and transposition (swap two characters in the string). The correction of a spelling error is based on probability theory where the most likely spell correction for a misspelt word is chosen based on the above-defined edit distance metrics.

So, we try to find a correction c, among all the possible candidate corrections to maximize the probability of c being the intended candidate, given the word w. Mathematically, it can be written as follows,

\begin{equation}
argmax_{c \in candidate} P(c|w)
\end{equation}

\subsection{Q-Map Algorithm and Architecture}

Q-Map is built as a very flexible system with various configurable options (All the examples presented here use UMLS Knowledge Source 2017AA). Each of these configurations are explained in detail later. These options help in defining how the system should parse the data for retrieving information. The Q-Map system can be broken down into three phases, namely, preprocessing, indexing, search. 

The speed of the system in search process can be attributed to the first stage, the preprocessing. All the overhead for maintaining minimal representation for achieving maximum overlap during search phase is dealt with in this step. The configuration setup during this step defines how aggressively the system annotates medical concepts in a clinical free text. Various configuration options available during this step include CUI clustering, stemming or lemmatization etc. The configuration options are explained below and the setup of the entire system is summarized briefly in Fig 1.

Limiting the number of semantic types can help in indexing only the relevant data depending on application requirement. For example, one can index data from only LOINC, RxNorm etc. if the retrieval aims at fetching drug information from the corpus. Also, many of the semantic types often have a lot of noise in terms of concepts present in them which might not be relevant to all types of applications. For example, the semantic type \qq{finding} has concepts named, \qq{disease}, \qq{syndrome} etc. which might be totally irrelevant when the application is medicine based because these concepts, even though have a medical association do not carry any importance in terms of information contained. Hence, depending on the use case, one can limit the number of semantic types used for indexing.

Removal of stop words while indexing is an option that can be toggled as per requirement. If the stopwords are removed, it can help in a successful match even when the inflection of terms in a document are different. For example, \qq{tumour of brain} and \qq{tumour in the brain} would match successfully if the stopwords are filtered. Similarly, other forms of the phrase, such as \qq{brain tumour} would also match successfully. This robustness can be attributed to the comprehensive collection of concepts in the Metathesaurus. It gives very promising results and does not require any additional effort to produce more forms of a given phrase.

Similarly, stemming is another option of preprocessing which helps in achieving uniformity when the terms are derivative of same root term. For example, \qq{diabetes} and \qq{diabetic} would both stem to \qq{diabet}. Even though the final word does not have a meaningful spelling sometimes, empirically the added benefits of uniformity far outweigh any drawbacks of lack in meaning due to spell errors. Alternatively, one can use lemmatization which is basically a dictionary lookup to get the accurate root word and would keep the spelling correct and meaning intact at the cost of increased processing time.

The preprocessing step standardizes the data in appropriate form for the second step, which is indexing to generate the FSM. Indexing consists of two steps in itself. The first step is the building of the trie where each node is a character. The step two is the main compute intensive step, where in for each node its failure state transition is determined that searches for other branches that have the longest prefix which is also the suffix of the failed branch. Also, the payload stored with each concept name in the index of FSM can be easily modified to store additional data which should be returned upon match of terms.

The speed of the algorithm can be attributed to this search automaton which reduces the time complexity of concept retrieval, which is proportional to the length of the document on which the algorithm is operating for mining the clinical concepts.

The third step of the Q-Map system is search, where same preprocessing configuration as used during the first step is applied on the corpus that is undergoing the annotation task in order to maintain the uniformity of analysis. 

The additional option of spell correction, introduced in section IV-C, is available for preprocessing the text before search phase.

There is a configurable option for the output generation from the search process. One can define whether the response should return all the concepts or the superset concepts only. By superset, we mean the string in the concepts mined that are not a substring of another string being returned. For example, if we search on a document containing \qq{Patient is diagnosed with type 2 diabetes mellitus and hypertension}, then the output of search on this document would return \qq{diabetes}, \qq{diabetes mellitus}, \qq{type 2 diabetes mellitus} and \qq{hypertension}. Here, \qq{diabetes} and \qq{diabetes mellitus} are the substrings of \qq{type 2 diabetes mellitus}. So, if the search is configured to return supersets only, then returned concepts would be only \qq{type 2 diabetes mellitus} and \qq{hypertension}. This option has various benefits when one wishes to keep the most informative terms only and not all the variations.

\begin{table*}
\label{table_example}
\caption{Comparison of Q-Map and MetaMap Outputs}
\begin{center}
\begin{tabular}{C{8cm}C{3cm}C{3cm}}
\hline \hline
Document & MetaMap Output & Q-Map Output \\
\hline
Ligament tear observed in Radiograph. Patient is a 53 year old female and has been recommended for unilateral Total Knee Replacement. Patient was given Heparin as a prophylactic treatment to prevent DVT. & \textbf{Diagnosis:} \newline
Ligament Tear \newline
\textbf{Procedures:} \newline
Total Knee Replacement \newline
Radiography \newline
Prophylactic Treatment \newline
\textbf{Medicines:} \newline
Heparin
 & \textbf{Diagnosis:} \newline
Ligament Tear \newline
\textbf{Procedures:} \newline
Total Knee Replacement \newline
Radiography \newline
Prophylactic Treatment \newline
\textbf{Medicines:} \newline
Heparin\\

\hline
Patient, F / u / c / o seizure disorder on CBZ, now admitted with nausea, vomiting and generalised  tiredness. EEG was normal. Dose of Tegretol was reduced. For throat discomfort and vomiting, ENT consultation and laryngoscopy was done and said to be normal. 
& 
\textbf{Diagnosis:} \newline
Seizure Disorder \newline
Vomiting \newline
Throat Discomfort \newline
\textbf{Procedures:} \newline
EEG \newline
Laryngoscopy \newline
\textbf{Medicines:} \newline
Tegretol 
& 
\textbf{Diagnosis:} \newline
Seizure Disorder \newline
Vomiting \newline
Throat Discomfort \newline
\textbf{Procedures:} \newline
EEG \newline
Laryngoscopy \newline
\textbf{Medicines:} \newline
Tegretol
\\

\hline
Patient is a 48 year old male and presented himself in the Emergency Department with chest pain. AFib and massive cardiac arrest was observed. ECG, Echocardiogram and Chest CT was ordered along with Troponin test, HBA1C test and Serum creatinine test. Patient was given Warfarin on a tapering down dosage starting with 3mg. Patient needs to undergo CABG. 
& 
\textbf{Diagnosis:} \newline
Chest Pain \newline
Cardiac Arrest \newline
\textbf{Procedures:} \newline
Chest CT \newline
ECG \newline
Serum Creatinine Test \newline
Echocardiogram \newline
\textbf{Medicines:} \newline
Warfarin 
& 
\textbf{Diagnosis:} \newline
AFib \newline
Cardiac Arrest \newline
Chest Pain \newline
\textbf{Procedures:} \newline
HBA1C Test \newline
Chest CT \newline
Troponin \newline
Echocardiogram \newline
Serum Creatinine Test \newline
CABG \newline
\textbf{Medicines:} \newline
Creatinine
\\
\hline \hline
\end{tabular}
\end{center}
\end{table*}

\subsection{Negation Detection}

The system implements negation detection based on the algorithm proposed in NegEx \cite{negex}. NegEx is a simple syntactic algorithm for identifying negations given the phrases to be tested for negations and the sentences they occur in. Based on the regular expression, the algorithm is capable of detecting different classes of negations, namely pseudo-negations and positive negations. Pseudo-negations consists of cases such as double negatives (e.g. not ruled out), modified meaning (e.g. gram-negative), or ambiguous phrasing (e.g. partial presence of symptoms).

With the help of NegEx, the system is capable of differentiating among positive and negative cases which is in-turn helpful in fixing false positives that would be detected otherwise.

\section {Results}

Both Q-Map and MetaMap are tools to retrieve medical concepts from free text datasets, but vary in their approaches and algorithms. MetaMap is based on prolog engine and has a few extra features like those of minimum phrase parsing of a sentence, word sense disambiguation (WSD). But we noticed that we did not require minimum phrase parser with our approach in Q-Map. This is mainly because of the rich data available in Metathesaurus which is used of the indexing and creating the Q-Map FSM.

Also, the performance of minimum phrase parser and WSD depends largely on how well the input text is structured. If the input is grammatically sound and follows the rules of sentence formation, then performance is appreciable. But it is observed that in many cases the sentence formation in clinical notes is not that well-structured. In either case there is no drop in performance of Q-Map because of the preprocessing steps done for data uniformity.

 Another major difference between the two systems can be seen in terms of the speed of system. MetaMap operates on data on disk while Q-Map is designed to operate in-memory which gives it a faster processing speed. There are a few drawbacks of this in-memory operations when the dataset being loaded is very large, but so far, we have created automatons with as many as 25 million nodes and have gotten maximum of 4GB size models, which are easily managed by the current machines.
 
Table I presents a few of real world discharge summaries and the medical concepts returned by both Q-Map and MetaMap.

In Table I, the concepts retrieved are grouped under the groups medicines, diagnosis and procedures for comparison purpose. This is done using the bucketing of sematic types associated with concepts and is easily available in Metathesaurus. From the table, it can be observed that the performance of Q-Map in retrieving concepts is on-par with that of the MetaMap. 

Particularly, in row 3 of Table I it can be noticed that Q-Map outperforms MetaMap in retrieving concepts related to medical procedures. This points to a small advantage it has over MetaMap because of digesting the Metathesaurus data after preprocessing.

Another advantage of Q-Map is that addition and debugging of processes in Q-Map is much easier because it can work with any form of Knowledge Source and applied in various different fields. The program size is comparably lesser than MetaMap and language of choice is python, which is more easily adaptable than the prolog engine integrated Java source code of MetaMap.

In terms of processing speed, Q-Map is found to be much faster and takes only around 22.9ms to process a 110 word discharge summary as against MetaMap which takes around 11.9s (experiments performed on a machine with Intel i7 processor and 32GB DDR4 RAM). Also, while it is possible to reduce memory usage by limiting semantic types in Q-Map, no such adjustments are possible in MetaMap.

\section{Conclusion and Future Scope}

In this work, we presented Q-Map, a system for clinical data retrieval and compared its performance with MetaMap. Empirical evidence shows that the system works as well as MetaMap, occasionally outperforming it. It also shows major improvement in terms of memory management by limiting semantic types and speed of processing documents by order of 500.

The added advantages of configurability, easy debugging, memory usage management and speed of processing make it a worthy substitute of the popular tool, MetaMap.
Experiments on texts from various sources have shown that Q-Map works very well for finding clinical concepts in human written free text. Q-Map has various utilities in the domain and is currently being used to build intelligent systems to predict correct ICD codes from the human written discharge summaries. It has also proved to be useful in creating features used to build decision trees using XGBoost and Random Forest. 

There are various improvements which can be incorporated to ensure better performance. Usage of deep learning models for concept aggregation would be more efficient than the current data driven approach used, but the unavailability of the data for training poses a challenge.

Q-Map also provides scope to augment the algorithm with semantic lookups in the text corpora along with the syntactic lookups which is implemented in the current version. Also, there is a scope to understand acronyms and abbreviations better. Though, the above enhancements could not be incorporated, Q-Map with its current version gives promising results and has advantages over its contemporary counterparts.



\ifCLASSOPTIONcaptionsoff
  \newpage
\fi

\begin{IEEEbiographynophoto}{Sheikh Shams Azam}
has a bachelor's degree in Electrical and Electronics Engineering from National Institute of Technology Karnataka (NITK) and is currently working as a Data Scientist at Enlightiks, Practo Technologies Pvt. Ltd. He is also the author of the blog Machine Learning Medium (site: https://machinelearningmedium.com/)
\end{IEEEbiographynophoto}

\begin{IEEEbiographynophoto}{Manoj Raju}
has a master's degree in Computer and Communications Technology from Saarland University (Germany) and is currently working as a Senior Data Scientist at Enlightiks, Practo Technologies Pvt. Ltd. He has about 5 years of experience working on cutting edge machine learning and deep learning technologies in computer vision, predictive and statistical modelling. 
\end{IEEEbiographynophoto}

\begin{IEEEbiographynophoto}{Venkatesh Pagidimarri}
 has over 9 years of experience in data analytics, statistical modelling, predictive modelling, machine learning and deep learning technologies. He has extensively worked in Healthcare space in developing patient centric risk models. Venkatesh holds master degree from IIT Madras.
\end{IEEEbiographynophoto}

\begin{IEEEbiographynophoto}{Vamsi Kasivajjala}
has over 17 years of experience creating multiple products predominantly in Healthcare Informatics and Analytics. As a part of his corporate journey, he has been successful in M\&A activities (3 acquisitions, and 2 Divestments). Been a P\&L owner and CTO responsible for growth and achieving targets YoY. He is the co-founder and CEO of Enlightiks and also the President and Board Member of HIMSS Asia Pacific India Chapter.
\end{IEEEbiographynophoto}

\end{document}